\documentclass[fleqn,usenatbib]{mnras}

\usepackage[T1]{fontenc}
\usepackage{ae,aecompl}

\usepackage{graphicx}	
\usepackage{amsmath}	
\usepackage{amssymb}	
\usepackage{threeparttable}
\usepackage{multirow}
\usepackage{url}
\usepackage{multicol}
\usepackage[dvipsnames]{xcolor}

\newcommand{\mum}{\ifmmode{\rm \mu m}\else{$\mu$m}\fi}

\newcommand{\Msun}{\ensuremath{{\rm M}_{\odot}}}      
    
\newcommand{\chisq}{\ifmmode{\chi^{2} }\else{$\chi^2$}\fi}
\newcommand{\rchisq}{\ifmmode{\chi^{2} }\else{$\chi^2_\nu$}\fi}

\newcommand{\spitzer}{{\em Spitzer }}

\title[Infrared Variable Stars in M32]{Infrared variable stars in the compact elliptical galaxy M32}

\author[O. C. Jones et al.]{O.~C.~Jones,$^{1}$\thanks{E-mail: olivia.jones@stfc.ac.uk}
C.~Nally,$^{2}$
M.~J.~Sharp,$^{2}$
I.~McDonald,$^{4,5}$
M.~L.~Boyer,$^{3}$
M.~Meixner,$^{3,9}$
\newauthor
F.~Kemper,$^{6,7}$
A.~M.~N.~Ferguson,$^{2}$
S.~R.~Goldman,$^{3}$ 
R.~M.~Rich$^{8}$
\\
$^{1}$UK Astronomy Technology Centre, Royal Observatory, Blackford Hill, Edinburgh, EH9 3HJ, UK\\\(\)
$^{2}$Institute for Astronomy, University of Edinburgh, Blackford Hill, Edinburgh EH9 3HJ, UK\\\(\)
$^{3}$Space Telescope Science Institute, 3700 San Martin Drive, Baltimore, MD 21218, USA\\
$^{4}$Jodrell Bank Centre for Astrophysics, School of Physics and Astronomy, University of Manchester, Oxford Road, Manchester M13 9PL, UK\\
$^{5}$The Open University, Walton Hall, Kents Hill, Milton Keynes, MK7 6AA, UK\\
$^{6}$European Southern Observatory, Karl-Schwarzschild-Str. 2, 85748 Garching b. M\"unchen, Germany \\
$^{7}$Academia Sinica, Institute of Astronomy and Astrophysics, Taipei 10617, Taiwan\\
$^{8}$Department of Physics \& Astronomy, University of California Los Angeles, Los Angeles, CA 90095, USA \\
$^{9}$SOFIA-USRA, NASA Ames Research Center, MS 232-12, Moffett Field, CA 94035, USA 
}

\date{Accepted XXX. Received YYY; in original form ZZZ}

\pubyear{2021}

\usepackage{newtxtext,newtxmath}

\begin{document}
\label{firstpage}
\pagerange{\pageref{firstpage}--\pageref{lastpage}}
\maketitle


\begin{abstract}
Variable stars in the compact elliptical galaxy M32 are identified, using three epochs of photometry from the {\em Spitzer Space Telescope} at 3.6 and 4.5 $\mu$m, separated by 32 to 381 days. We present a high-fidelity catalogue of sources detected in multiple epochs at both 3.6 and 4.5 $\mu$m, which we analysed for stellar variability using a joint probability error-weighted flux difference.
Of these, 83 stars are identified as candidate large-amplitude, long-period variables, with 28 considered high-confidence variables. 
The majority of the variable stars are classified as asymptotic giant branch star candidates using colour-magnitude diagrams. 
We find no evidence supporting a younger, infrared-bright stellar population in our M32 field. 
\end{abstract}


\begin{keywords}
galaxies: individual (M32) --- infrared: galaxies --- infrared: stars -- stars: late-type -- galaxies: stellar content ---  stars: variables
\end{keywords}



\section{Introduction}
\label{sec:intro}

M32 is an inner satellite galaxy of the Andromeda spiral galaxy (M31), and is our nearest \citep[785 kpc;][]{McConnachie2005} example of a compact elliptical galaxy. These rare galaxies have very high stellar densities, small effective radii ($r_{\rm{eff}} \sim 0.1-0.7$ kpc) and luminosities of $\sim 10^9 \, {\rm L}_{\odot}$ \citep{Graham2013}. Such galaxies are thought to have formed via the tidal stripping of larger galaxies \citep[e.g.][]{Faber1973,Bekki2001,FerreMateu2018} or intrinsically as low-mass `early-type' galaxies \citep{Kormendy2009,Martinovic2017}.
Indeed, M32 has been recently implicated to be the remnant core of a galaxy that underwent a significant merger with M31 \citep{DSouza2018}.

M32 has had a prolonged star formation history, with a large intermediate-age ($\sim$2--8 Gyr) population spanning a range of metallicities; the peak occurring at [Fe/H] $\sim -0.2$ \citep{Grillmair1996,Monachesi2012,Davidge2014,Jones2015a} and a centrally concentrated, young ($<1$ Gyr), metal-rich ([Fe/H]$\sim +0.1$) stellar population \citep{Trager2000b, Rose2005,Coelho2009}.
To date, the analysis of M32's variable star population has focused on RR Lyr variables \citep{Fiorentino2012,Sarajedini2012}, which are indicative of an ancient ($>10$ Gyr) population, found to be uniformly mixed across M32. 
It has also been estimated that $\sim$60\% of the brightest asymptotic giant branch (AGB) stars in M32 are long-period variables \citep[LPVs;][]{Davidge2004}. Moreover, OGLE results suggest that all AGB stars are LPVs \citep[e.g.][]{Soszynski2013}, including the most extreme dust enshrouded AGB stars \citep[e.g.][]{Wood1992,Whitelock2003}.


Thermally pulsating AGB stars (TP-AGB) are often long-period, large-amplitude variables \citep{Iben1983, Wood1998, Ita2004, Whitelock2017}, and can have mid-IR excess emission due to warm, circumstellar dust as illustrated by the {\emph Spitzer} SAGE survey \citep[e.g.][]{Blum2006,Boyer2011,Matsuura2009,Srinivasan2016}. LPVs pulsate with periods of $\sim$60--1000 days \citep{Vassiliadis1993}, and can be classified as Mira, semi-regular, or irregular variables \citep{Fraser2008,Soszynski2009, Trabucchi2017}.
In the most extreme cases, AGB stars may vary on timescales longer than 300 days; these stars experience intense mass-loss rates (from $10^{-6}$ to $10^{-4}$ M$_{\odot}$ yr$^{-1}$; \citealt{Vassiliadis1993}) and are important contributors to the chemical enrichment of the interstellar medium (ISM).


TP-AGB stars are highly luminous, particularly in the infrared, and are excellent tools for studying the resolved stellar populations in nearby galaxies, particularly those with older populations, or where there is substantial interstellar obscuration \citep[e.g][]{Menzies2002,Whitelock2009,Whitelock2013,Menzies2015}. LPVs have been used to infer the star formation histories (SFHs) of nearby  galaxies \citep{Javadi2011, Rezaeikh2014, HamedaniGolshan2017, Hashemi2019, Navabi2021}. They may also rival Cepheid variables as fundamental calibrators of extragalactic distances \citep{Whitelock2012, Huang2020}. 
%

Previous {\em Spitzer} variability studies of Local Group galaxies have obtained between two and eight epochs of imaging, and were able to identify a large population of dust-producing AGB star candidates with 3.6-$\mu$m amplitudes up to 2.0 mag \citep{LeBertre1992,McQuinn2007,Vijh2009,Riebel2010,Polsdofer2015, Boyer2015b, Goldman2019, Karambelkar2019}. 
In this paper, we investigate for the first time in the mid-infrared (IR) the variable stellar population of the compact elliptical galaxy M32 with {\em Spitzer}. The photometric observations and data reduction are discussed in Section~\ref{sec:observations}, in Section~\ref{sec:results} we identify variable stars and determine their stellar classifications. Our conclusions are summarised in Section~\ref{sec:conclusion}.

\section{Observations and Photometry}
\label{sec:observations}

\begin{figure}
\centering
\includegraphics[angle=-0.5, trim=0.2cm 0.2cm 0.1cm 0cm, clip=true,width=\columnwidth]{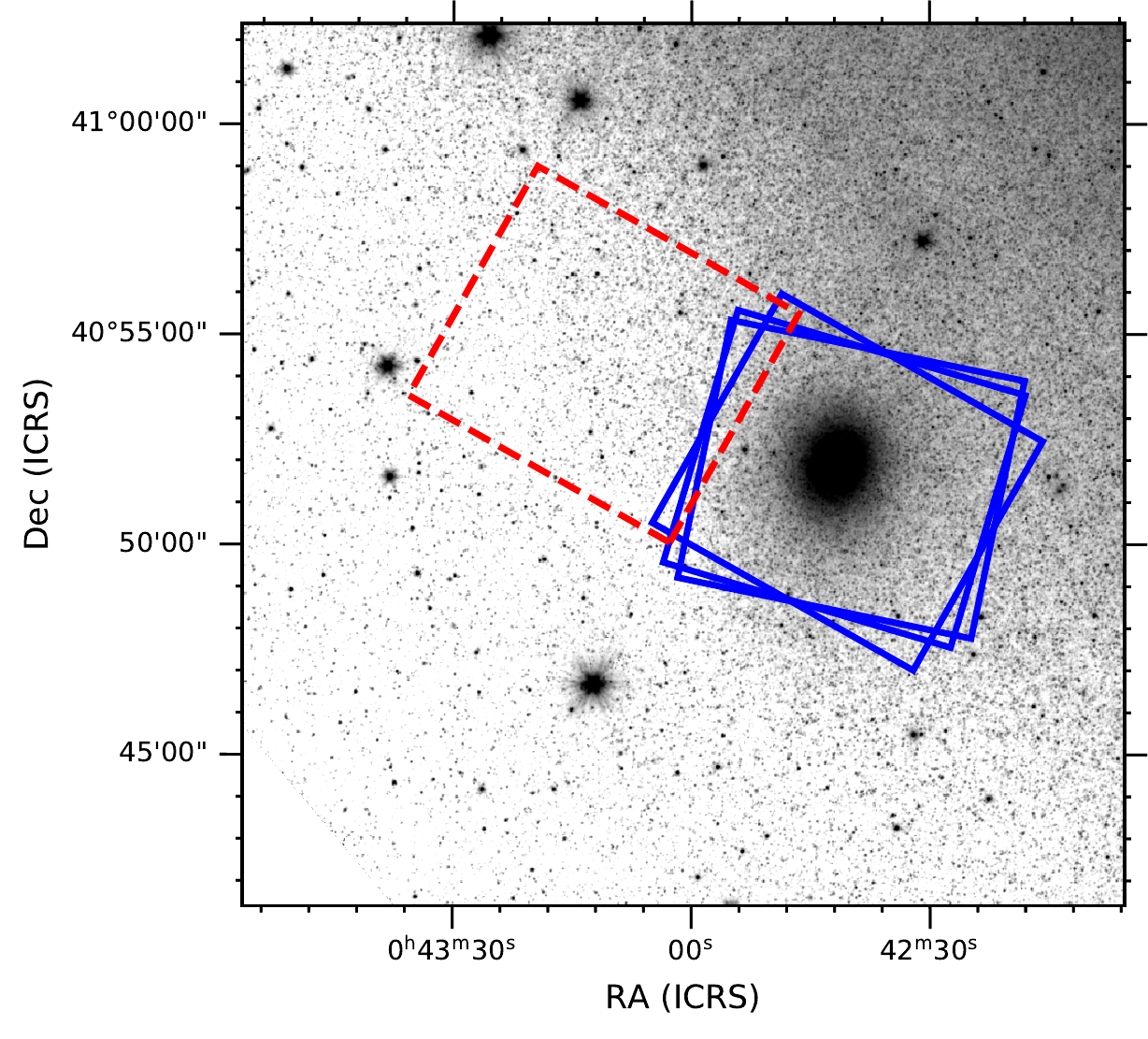}
 \caption{Location of our {\em Spitzer} IRAC pointings towards M32 (blue solid lines) superimposed on a IRAC 3.6 $\mu$m mosaic of M31 \citep{Barmby2006, RafieiRavandi2016}. We limit our variability search to the region centred on M32, where these pointings and the IRAC 3.6 and 4.5 $\mu$m data overlap. The red dashed line indicates the location of the M31 field.} 
  \label{fig:M32_obs_cov}
  \end{figure}

Observations of M32 (program ID 11103) were made with the Infrared Array Camera \citep[IRAC;][]{Fazio2004} on the {\em Spitzer Space Telescope} using the 3.6 and 4.5 $\mu$m filters \citep{Werner2004}. The three epochs were taken over a 13 month period (Table~\ref{tab:obs_summary}) during the post-cryogenic mission. 
Each pointing consists of 23 dithered exposures of 30s in each IRAC filter.
The region where both 3.6 and 4.5 $\mu$m data are available covers an area of approximately 5$^{\prime}$ $\times$ 5$^{\prime}$ around the centre of M32, with a pixel size of 0.6$^{\prime\prime}$. This corresponds to a 1.3 kpc$^2$ field-of-view. 
As M32 is projected against the disc of M31, we have also obtained a single background field (imaged to the same depth), at a similar isophotal radius in M31, to establish contamination statistics for our sample. As this M31 field is located slightly further from M31 than our primary M32 observations, the contamination from M31 will likely be underestimated in our corrected M32 star counts (see Section~\ref{sec:contaminationCorr}). The locations of both fields are shown in Figure~\ref{fig:M32_obs_cov} and their cadence given in Table~\ref{tab:obs_summary}.

The raw IRAC data were processed by the Spitzer Science Center reduction pipeline version 19.2.0
and were further reduced with the {\sc mopex} data reduction package \citep[][]{MakovozMarleau2005}. This helps correct for imaging artefacts such as stray light, column pulldown and bad pixels.
%
%
%
Point-spread function (PSF) photometry was performed on the co-added, mosaicked data from all epochs, separately, to achieve the deepest photometry possible. 
PSFs were generated using at least 20 bright, isolated stars in each IRAC band and epoch, and sources with a $5\sigma$ detection above the local background were chosen for extraction. The PSF photometry was conducted using the {\sc daophot ii} and {\sc allstar} photometry packages \citep{Stetson1987}, which are optimised for crowded fields.
We implement strict point-source detection criteria by adopting a sharp and round cut-off for all sources detected at 3.6 and 4.5 $\mu$m (only sources with sharpness and roundness values within 1.75$\sigma$ of the respective mean values were kept); this effectively removes contamination from cosmic rays, stellar blends and minimizes the number of extended sources in the catalogue.

%
%

As recommended by the {\em Spitzer} Science Centre\footnote{\url{https://irsa.ipac.caltech.edu/data/SPITZER/docs/irac/}}, 
we apply a colour-correction to the flux densities using a 5000~K blackbody, typical of a red-giant branch (RGB) star at the base of the RGB \citep{McDonald2017}. An array-location-dependent correction\footnote{\url{https://irsa.ipac.caltech.edu/data/SPITZER/docs/irac/calibrationfiles/locationcolor/}} was applied to sources with $[3.6]-[4.5] < 0$ mag to correct for variations in point source flux across the array \citep{Quijada2004} due to flat-fielding, and a pixel-phase-dependent correction \citep{Reach2005} was applied to the 3.6-$\mu$m photometry to correct for quantum-efficiency variations across pixels. Finally, magnitudes relative to Vega were derived using a zero-magnitude flux of 280.9 $\pm$ 4.1 Jy for 3.6 $\mu$m and 179.7 $\pm$ 2.6 Jy for 4.5 $\mu$m, as specified by the {\em Spitzer} IRAC Data Handbook version 2.1.2\footnote{See footnote 1.}. 
 
Figure~\ref{fig:completeLim} shows the representative photometric uncertainty as a function of source magnitude. Photometric errors include standard {\sc daophot ii} errors and the IRAC absolute calibration errors of 3 per cent \citep{Reach2005}. For stars included in the final catalogues, the median photometric uncertainty is 5.5 per cent. These have not been adjusted to account for foreground interstellar extinction.
 
%
%
\subsection{Photometric completeness and stellar crowding}
 \label{sec:Completeness}
 
\begin{figure} 
\centering
\includegraphics[trim=0.5cm 0cm 0.5cm 0cm, clip=true,width=\columnwidth]{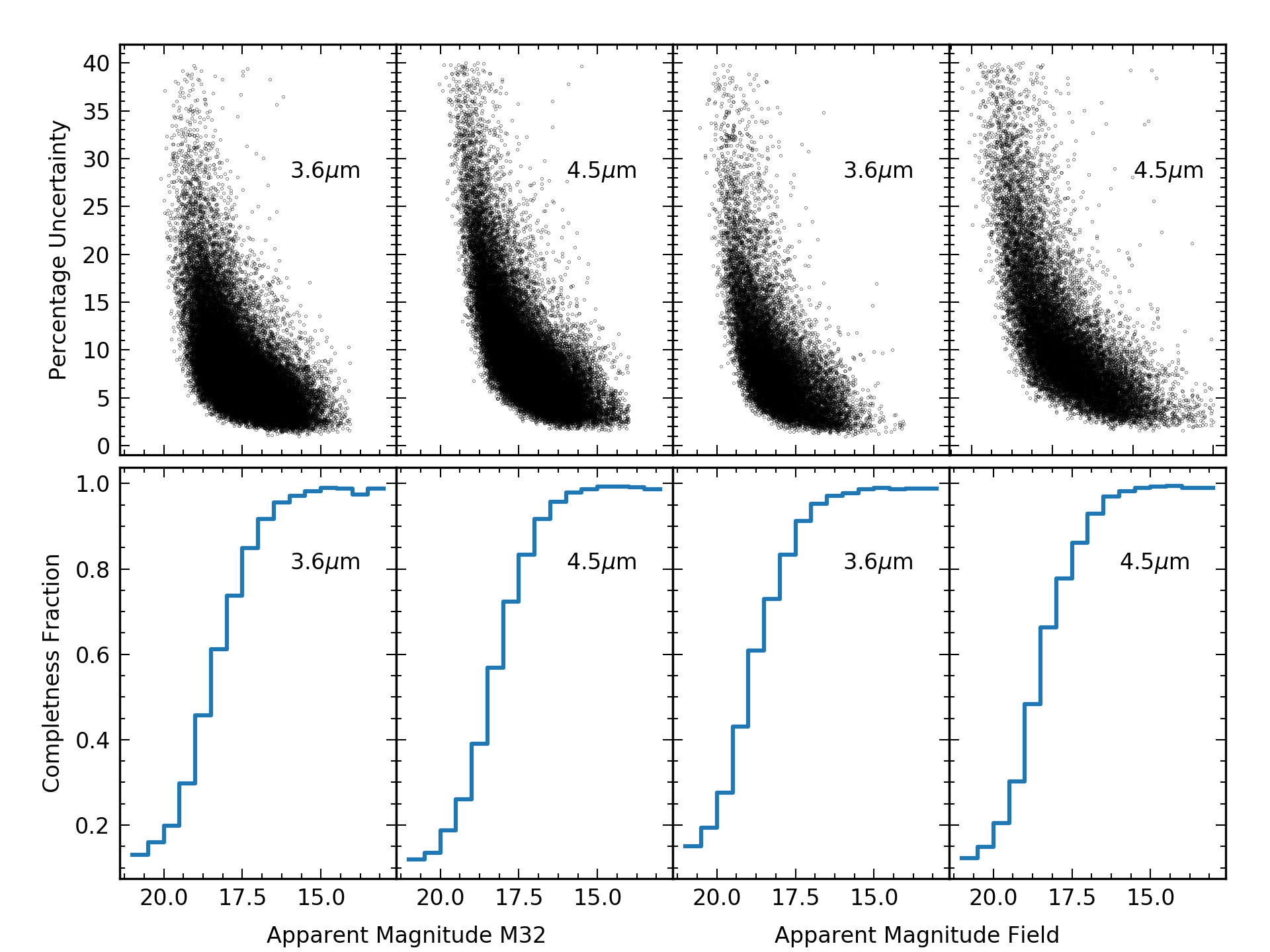}
\caption[Photometric Completeness]{Photometric uncertainty and completeness fraction as a function of apparent magnitude for each IRAC band, for both the M32 pointings (left) and the M31 field (right).}
 \label{fig:completeLim}
\end{figure}

Artificial star tests were used to determine the completeness limits of our sample. False stars were injected at random pixel locations (excluding the galaxy centre; $R \lesssim  0.5$\arcmin, due to the severe crowding) across the image, with a limiting magnitude $\sim$2 mag fainter than the extracted photometric catalogue.
Sources are considered to be recovered if they are within a one-pixel radius of the input position and their magnitude differs by $|\delta m| \leq 1$ mag from their input magnitudes. 
The magnitude limit ensures that we recover reasonably accurate magnitudes and prevents the recovery of sources that are products of blends \citep[e.g.][]{Monachesi2011, Jones2015a}.
This process was repeated 100 times per field/filter combination, each time injecting only a small number of false stars per iteration ($<$5 per cent of the original stars observed), to avoid increasing the crowding in the images.
The average completeness curves, comparing the fraction of injected to recovered stars for each IRAC band, are shown in lower panel in Figure~\ref{fig:completeLim}.
These tests indicate that the catalogue is 80\% complete to 17.22 mag at 3.6 $\mu$m and 17.15 mag at 4.5 $\mu$m for M32 and complete to 17.66 mag at 3.6 $\mu$m and 17.36 mag at 4.5 $\mu$m for the M31 field stars (the M31 observations are more sensitive due to experiencing less photometric crowding).

\begin{figure}
\includegraphics[trim=1cm 0cm 0cm 1cm, clip=true,width=1.1\columnwidth]{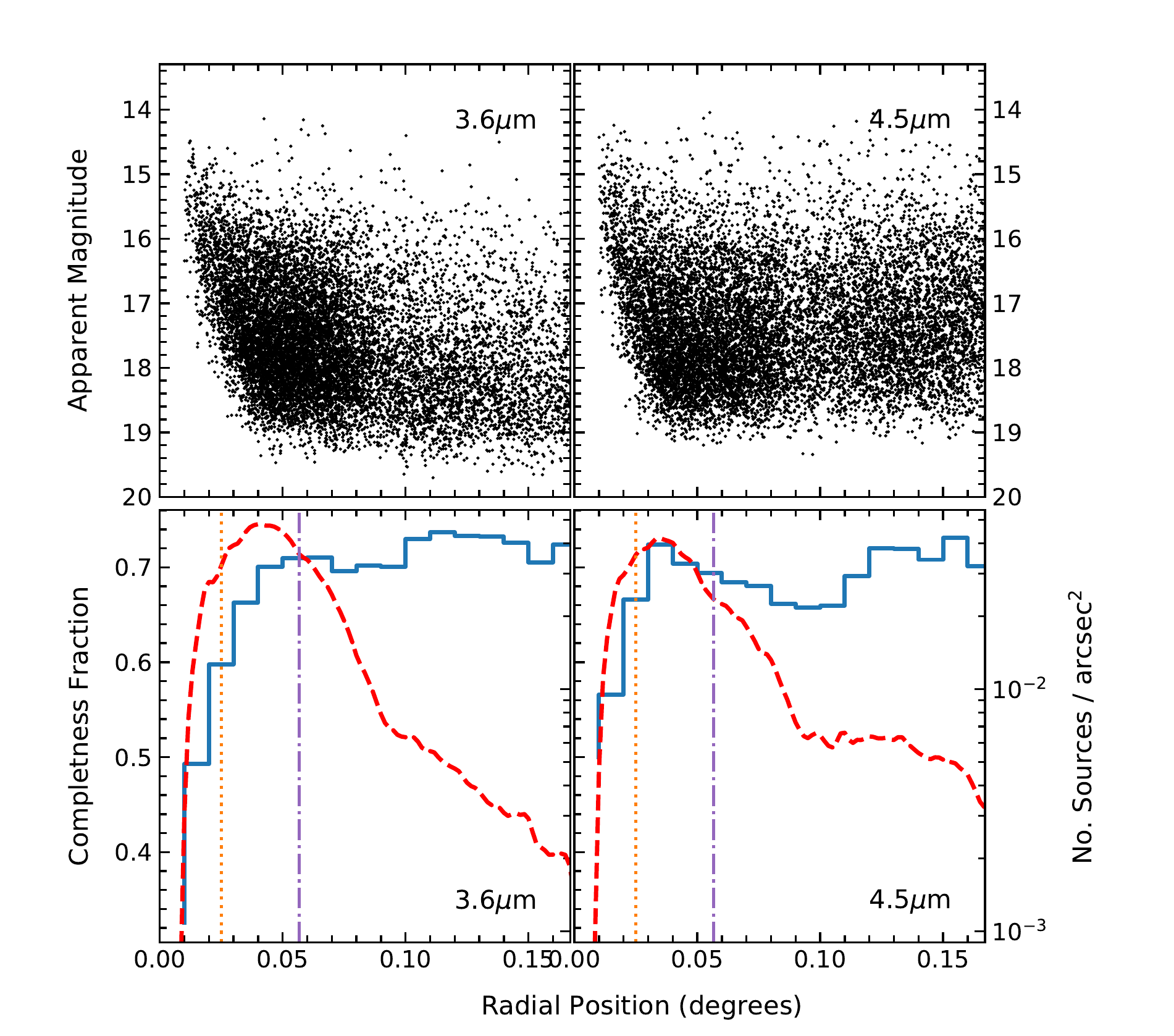}
 \caption[Apparent magnitude versus radial distance]{Distribution of the apparent magnitude (top) and 
 photometric completeness fraction (bottom, blue line) with radial distance. The red dashed line indicates the number of sources per arcmin$^2$. Towards the nucleus of M32 crowding is severe; the very bright sources at small radii likely receive a substantial magnitude enhancement from other unresolved, blended stars.  The orange dotted line marks the core region of M32; the dashed-dot purple line indicates the region where M31 begins to contribute substantially to the source density.}
  \label{fig:Blendcheck}
\end{figure}

Completeness is a function of the crowding level; given the wide range of  projected stellar densities found across the field of view (FOV) we examine, in Figure~\ref{fig:Blendcheck}, the photometric completeness and the apparent magnitude of the population as a function of radius from the centre of M32. In the inner regions of M32 ($R \lesssim 1.5^\prime$), the completeness fraction drops rapidly and the crowding is so severe that  no individual stars can be reliably resolved. 
Given M32's steep brightness profile and the high degree of crowding and blending towards the core, we exclude the galaxy centre and inner regions ($R \lesssim 1.5^\prime$) from our analysis. 
 As the optical half-light radius of M32 is 0.47$^\prime$ \citep{McConnachie2012}, our resolved sources constitute only a small fraction ($\sim20\%$) of the total stellar mass of M32 if we assume a de Vaucouleurs profile.
Thus the number of long-period variables identified in M32 should be considered a lower limit.

 Slightly further out from the severely crowded centre ($R =  1.5^\prime$ to $2.4^\prime$), some stars are resolvable but the galaxy surface brightness is still high. Bright sources may exhibit a magnitude enhancement as they are photometrically inseparable from superimposed blended sources, while faint sources are less effectively recovered. 
The variable surface-brightness levels and significant substructure in the M31 outer disc is reflected by the undulating profile of the completeness fraction at large radii from M32's centre. 
M32 has a stellar density distribution which falls off rapidly from its core \citep{Lauer1998,Graham2002,Jones2015a}, and is not expected to have a substantial halo population out to large radii. This is especially problematic for the [4.5] data, which is oriented towards the disc of M31 and includes sections of its spiral arms.
We consider stars with $R \lesssim 3.4^\prime$ to be probable M32 members. Beyond this, field stars from M31 begin to dominate the source density. This agrees well with \cite{Jarrett2019} who measured the semi-major axis of M32 in the \emph{WISE} Extended Source Catalogue to be 3.62$^\prime$ at 3.4 $\mu$m, and 2.95$^\prime$ at 4.6 $\mu$m.


%
%


\subsection{Description of the catalogue}

The individual point-source lists (three epochs at 3.6 $\mu$m and 4.5 $\mu$m) for the regions where both 3.6 and 4.5 $\mu$m data are available are cross-matched using a 1$^{\prime\prime}$ radius via the Bayesian-based code {\sc nway} \citep{Salvato2018}, producing a high-fidelity catalogue of over 1000 M32 sources detected in multiple epochs at both 3.6 and 4.5 $\mu$m, which is described in Table~\ref{tab:M32catDescription}. Sources were included in the catalogue of good sources if $p_{\rm any}>0.5$ (the probability a photometric point has a counterpart) and $p_{\rm i} > 0.8$ (the probability the given match is the correct counterpart).
If there are multiple matches that meet these criteria, then we select the most probable counterpart. The photometric catalogue is available on-line through VizieR, and contains only sources determined to be highly reliable, with well constrained errors  ($\delta m_{3.6}<$ 0.17 mag in all epochs), and which are not saturated, probable blends or present in only one band or epoch. We have not corrected the photometric catalogue for extinction as foreground extinction (E($B-V$) = 0.08 mag) is low towards M32, especially in the IR.

\begin{table*}
\centering
\caption{Journal of observations.}
\label{tab:obs_summary}
\begin{tabular}{@{}cccccc@{}}
\hline
\hline
Field/Epoch & UT Date &  AOR No. & RA. & Dec.  \\
\hline
M32\_E1	&	2015/03/19	&		53090048	&	0h42m41.83s	&	$+$40d51m55.0s	\\	
M32\_E2	&	2015/04/20	&	    53090304	&	0h42m41.83s	&	$+$40d51m55.0s	\\	  
M32\_E3	&	2016/04/03	&		53090816	&	0h42m41.83s	&	$+$40d51m55.0s	\\	
\hline
M31 Field	&	2015/04/20	&	53089792	&	0h43m07.89s	&	$+$40d54m14.5s	\\	
\hline
\end{tabular}
\end{table*}

\begin{table*}
\centering
\caption{Description of the catalogue of point sources for M32.} 
\label{tab:M32catDescription}
\begin{tabular}{@{}lllc@{}}
\hline
\hline
Column &    Name        & Description                                         & Null \\
\hline
1       &     ID        &  Point-source name                                  & \ldots \\
2       &     RA        &  Right ascension (deg) ; (J2000)                                & \ldots \\
3       &     Dec       &  Declination (deg) ; (J2000)             & \ldots \\
4--9    &  mag\_36\_N   &  3.6~\micron\ mag and uncertainty for Epochs 1--3   & -99.0 \\
10--15  &  mag\_45\_N   &  4.5~\micron\ mag and uncertainty for Epochs 1--3    & -99.0 \\
16--18  &  var\_36\_N   &  Variability index at [3.6] for interval N          & -99.0 \\
19--21  &  var\_45\_N   &  Variability index at [4.5] for interval N          & -99.0 \\
22      &  mean\_36     &  Mean 3.6~\micron\ mag  ($\langle m_{3.6}\rangle$)  & \ldots \\
23      &  mean\_45     &  Mean 4.5~\micron\ mag  ($\langle m_{4.5}\rangle$)  & \ldots \\
24      &  amp\_36      &  Observed [3.6] Amplitude (mag) ($\Delta m_{3.6}$)  & \ldots \\
25      &  amp\_45      &  Observed [4.5] Amplitude (mag) ($\Delta m_{4.5}$)  & \ldots \\
26      &  var          &  Variable star classification; candidate variable = 1; high-confidence variable = 2     & \ldots \\
\hline
\end{tabular}
\end{table*}

\section{Results}
\label{sec:results}

\subsection{Contamination from M31 and foreground Stars}
\label{sec:contaminationCorr}


M32 has a projected separation of 24$^{\prime}$ (5.4 kpc) from M31's centre and is seen against the disc of M31.  A background field in M31's disc located at a slightly fainter isophotal radius in M31's disc
then M32 (see Table \ref{tab:obs_summary}) was observed to statistically estimate the level of contamination from M31's stellar population.
We only consider the region covered by both the 3.6 and 4.5 $\mu$m data in the contamination estimates, as they should contain statistically similar populations of stars from M31, and both filters are required to place sources in the colour-magnitude diagram (CMD).
We can therefore statistically correct for contamination by subtracting stars with similar colours and magnitudes to the population observed in the control field.  However, given the difference in stellar density between the fields, crowding is more significant in M32 and the completeness fraction of the two samples needs to be accounted for. 
We perform the statistical decontamination following the procedures of \citet{Gallart1996,Monachesi2011}. Namely, the CMDs of the M32 and M31 fields are split into a series of magnitude bins. For each bin of 0.5 magnitudes, a number of stars ($F_n$) is identified, where
\begin{equation} 
F_n= \frac{\Lambda_{i}^{\mathrm{F\_M32}}}{\Lambda_{i}^{\mathrm{F\_M31}}} \, C_{i}^{\mathrm{F\_M31}}.
\end{equation}  
Here, $\Lambda_{i}^{\mathrm{F\_M32}}$ and $\Lambda_{i}^{\mathrm{F\_M31}}$ are the completeness fractions in bin $i$ of the CMDs of each galaxy, as calculated in Section~\ref{sec:Completeness}, and $C_{i}^{\mathrm{F\_M31}}$ is the number of sources in bin $i$ of the M31
CMD.  For each bin, these $F_n$ stars are removed at random from the M32 field. Nominally both pointings cover the same size area of sky, however the high stellar density in the core of M32 effectively reduces the area where we can recover the resolved stellar population of M32 to approximately half that of the background field; we account for this difference in the effective FOV when applying the statistical correction. 
After statistically correcting for contamination, 659 stars belonging to M32 remain.
For the remainder of this paper, when considering the properties of M32's stellar population, stars that we {\em statistically} consider M31 contaminants are plotted as grey diamonds and the remaining M32 population as black points.


This statistical removal of the M31 field should also statistically remove foreground (Galactic) stars and background galaxies. An estimate of the foreground star contamination may be obtained by comparison with the {\sc trilegal} stellar population synthesis code of \cite{Girardi2005}. Foreground sources are simulated for a 25 arcmin$^2$ field centred on M32. In this region we expect approximately 75 foreground stars brighter than 18.2 mag in our sample, with [3.6]--[4.5] colours of approximately zero. 
At the distance of M32 (785 $\pm$ 25 kpc; \citealt{McConnachie2005}), it is difficult to separate dusty stars with $[3.6]-[4.5] > 0.2$ mag from unresolved background galaxies \citep{Mauduit2012, Kozlowski2016}. 
To estimate the number of potential background galaxy contaminants, we use the \emph{Spitzer} Extragalactic Representative Volume Survey \citep{Mauduit2012} and the decadal mid-IR variability Survey of the B\"ootes Field \citep{Kozlowski2016}. From these surveys, we estimate there to be $<$50 background galaxies in our FOV brighter than 18.2 mag at 3.6 $\mu$m.
While active galaxies can vary, they do so irregularly (so are less likely to be detected in three epochs) and strongly variable galaxies are not as common as AGB variables generally. So when considering only variables, an AGB nature is more likely.
Furthermore, due to the high stellar density and surface brightness of M32, and given the initial photometric quality cuts to remove non-point-like sources, we expect the number of galaxies in our raw source counts to be lower than this.
Thus it is unlikely that such galaxies pose significant levels of contamination. We consider their contribution as negligible when considering only variable sources \citep{Polsdofer2015, Kozlowski2016}. 

\subsection{Luminosity functions}

\begin{figure} 
\centering
\includegraphics[trim=0cm 0cm 0cm 0cm, clip=true,width=\columnwidth]{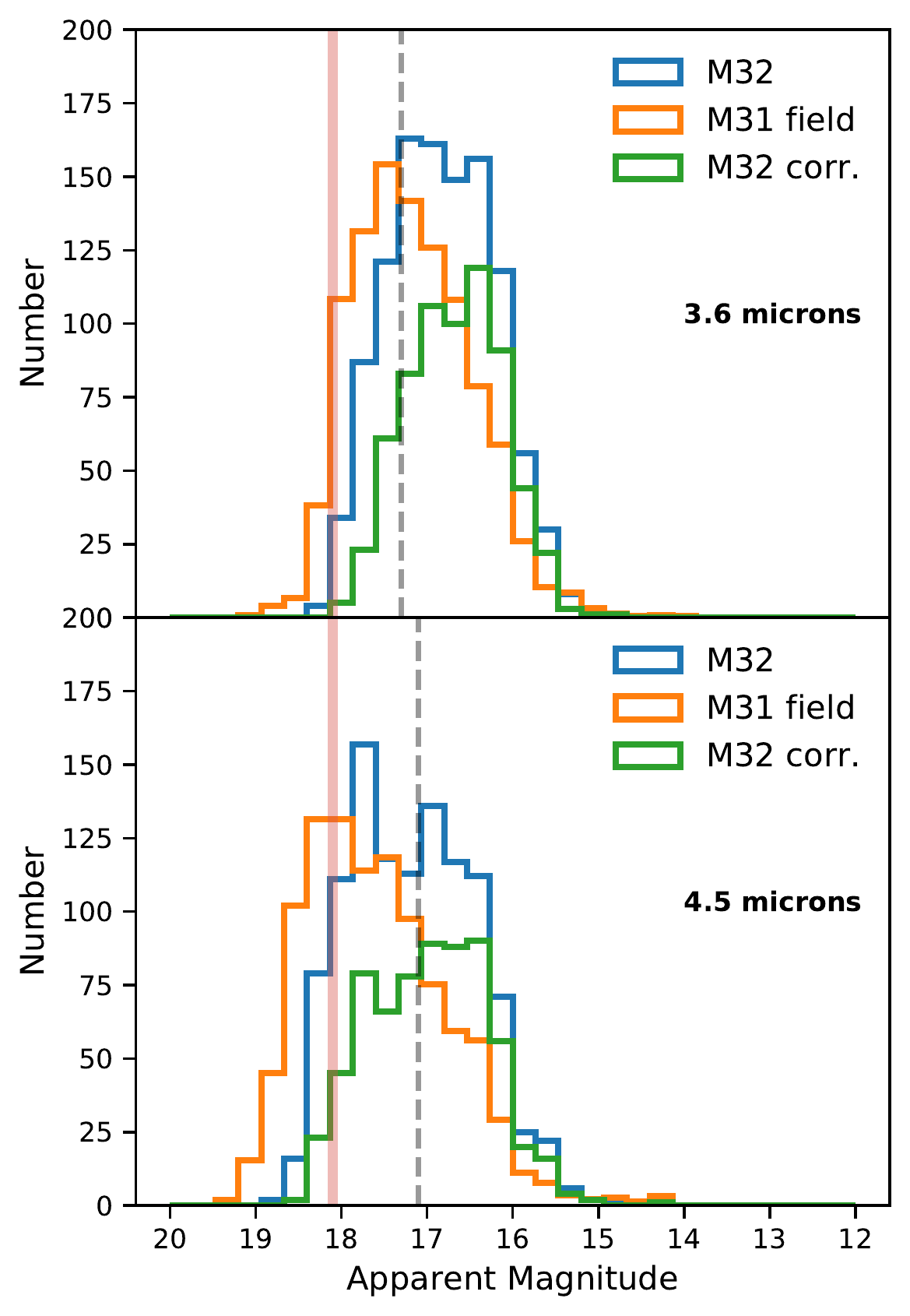}
\caption[Comparison of the M32 luminosity functions]{Comparison of the M32 and M31 (field) luminosity functions for 3.6 and 4.5$\mu$m. The contamination-corrected M32 luminosity function is shown in green. The vertical, dashed line marks the 80\% completeness limit for the M32 pointing, and the pale red line marks the expected tip of the RGB.} 
  \label{fig:M32_LumFunction}
\end{figure}

In Figure~\ref{fig:M32_LumFunction}, we plot the mean 3.6 and 4.5 $\mu$m luminosity functions for all the stars detected in the M32 and M31 fields.  The M31 star counts are scaled to the same effective survey area of the M32 pointing where its stellar populations can be comprehensively resolved.
The M32 luminosity function, statistically corrected for contamination from M31 stars is also shown. 
Although the data for the M31 field reach fainter magnitudes than the M32 data (due to the lower crowding) our analysis in both cases is limited to the brightest stars.
Evolved stars on the RGB, AGB and red supergiants (RSGs) are some of the brightest objects in the infrared sky at these wavelengths. 
At the distance of M32, the tip of the red-giant branch (TRGB) is expected at $m_{3.6} \sim 18.2$ $\pm$ 0.5 mag \citep{Davidge2014, Jones2015a}, thus our sample is biased toward discovering AGB (rather than RGB) variables: RGB stars populate an area of the luminosity function well below the completeness limit of our survey. 
 RGB amplitudes in the IR are also lower than AGB stars \citep{Boyer2015b} and are below the variability detection threshold of our data (see Sec.~\ref{sec:Var}).
In contrast, the lack of recent star formation during the last 50 Myr in M32 \citep{Brown1998} precludes moderately young stars like luminous RSGs from belonging to the galaxy; we thus assume that the majority of stars in our catalogue are likely to be intermediate-age AGB stars.

The M32 luminosity function at $3.6\mu$m peaks at brighter magnitudes compared to the M31 field stars, although if this is also true for the $4.5\mu$m data is unclear due to our photometric completeness limits.
Possible explanations for this include: (1) M32 has a younger population than M31's disc. (2) M32 might have undergone a period of enhanced star-formation several Gyr ago, and this history is now exhibited by its AGB population:
this possibility is intriguing as \cite{DSouza2018} postulate that M32 is the remnant core of a large spiral system which started interacting with M31 roughly 5 Gyr ago. (3) A difference in metallicity between the populations may affect the brightness of the AGB star populations, for instance the fraction of carbon stars would affect the dust emissivity at 3.6 and 4.5 $\mu$m. (4) M32 lies significantly in front of M31. (5) Our $3.6\mu$m data might be affected by stellar blends and crowding.

The position of M32 with respect to M31 has been subject of numerous studies. It is currently thought M32 is located in front of the M31 disk \citep[e.g][]{Ford1978, Choi2002, Georgiev2015}, but not sufficiently to explain the observed difference in the luminosity function peak.  
Stellar blending is a more likely possibility. For M32 \citet[][Section 2]{Jones2015a} noted that magnitude enhancements on IRAC point-source data are a significant problem for the inner regions $R < 1.5^{\prime}$ of M32, due to blends and Eddington bias, but magnitude enhanced sources form only a minor component of our catalogue outside this region. Our data exhibits a similar behaviour and so will only have a minor effect on the luminosity function.
We explore differences between the populations in Section~\ref{sec:M32cmds}, but do not consider there to be sufficient evidence to decide between these explanations here.

\subsection{Variability search}
\label{sec:Var}

\begin{figure} 
\centering
\includegraphics[trim=0cm 4cm 0cm 4cm, clip=true,width=0.8\columnwidth]{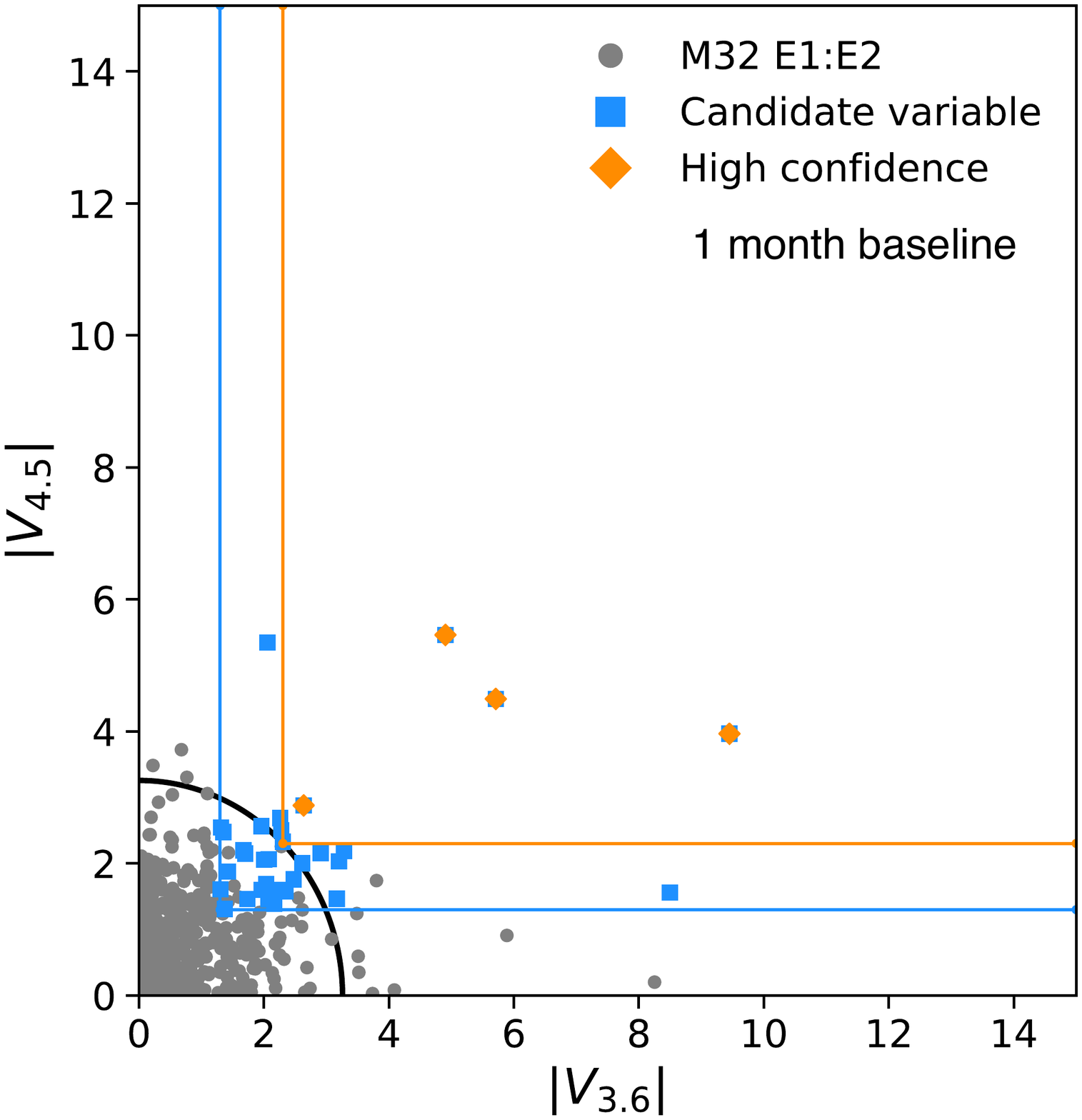}
\includegraphics[trim=0cm 4cm 0cm 4cm, clip=true,width=0.8\columnwidth]{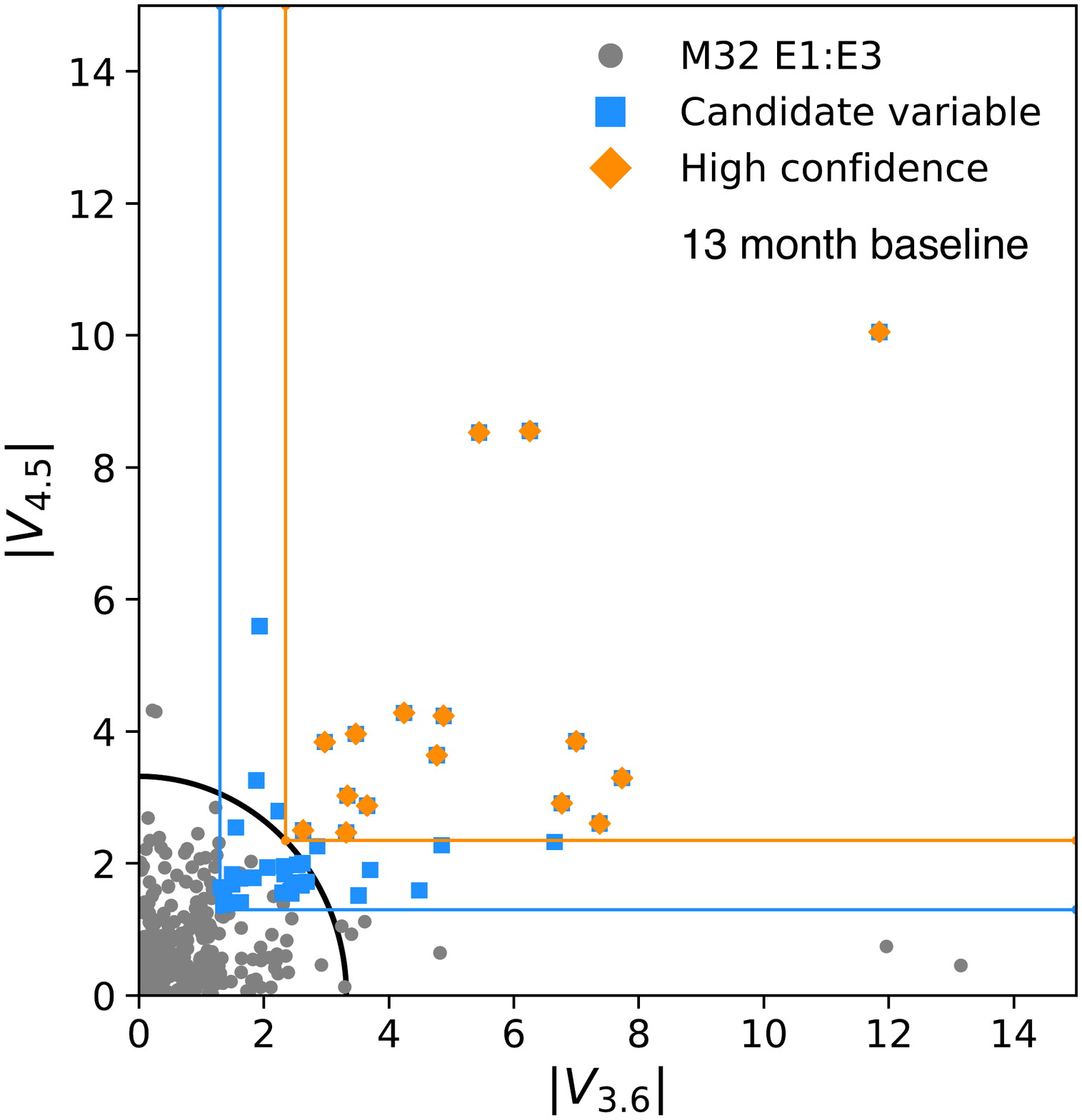}
\includegraphics[trim=0cm 4cm 0cm 4cm, clip=true,width=0.8\columnwidth]{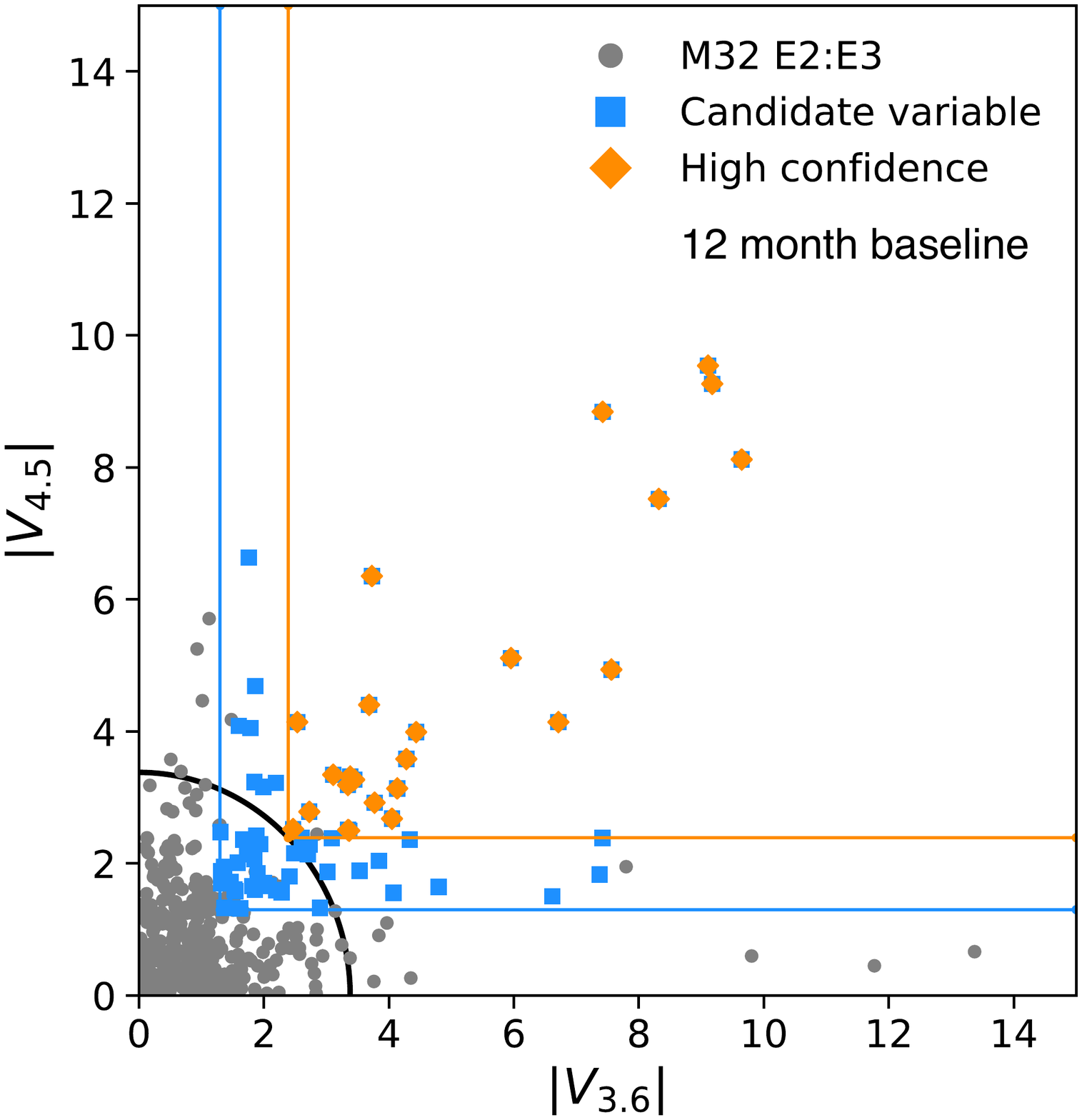}
\caption{The absolute values of the variability indices ($|V|$) at 3.6 and 4.5 $\mu$m for sources, from photometry in epochs 1 and 2 (top), 1 and 3 (middle), and 2 and 3 (bottom). The black arc indicates the 3$\sigma$ joint probability value assuming a 2D Gaussian distribution for $V_{3.6\mu{\rm m}}$ and $V_{4.5\mu{\rm m}}$. Orange points denote the high-confidence variables and blue squares denote candidate variables in each time interval.}
  \label{fig:M32_varCalc}
\end{figure}

The search for variable stars in sparse data sets that have systematic and measurement errors is a complex problem. To overcome data limitations there are a vast array of detection strategies (e.g.~direct image comparison, variability indices and periodicity search) that may be utilised to identify these interesting sources depending on the data coverage \citep[see][for a comprehensive comparison between methods]{Sokolovsky2017}.
Here, candidate variable stars are identified using the error-weighted flux difference between each pair of epochs. This procedure outlined by \citet{Vijh2009} for the Large Magellanic Cloud, and has been applied to other galaxies in the Local Group by \citet{Polsdofer2015,Boyer2015b,Jones2018a}  and \citet{Goldman2019}.
For each pair of epochs in a photometric band, a variability index $V$ is computed:

\begin{equation}
V = \frac{f_i - f_j}{\sqrt{\sigma_{f_i}^2 + \sigma_{f_j}^2}}
\end{equation}

\noindent
where $f_i$ and $f_j$ are fluxes for a source in epochs $i$ and $j$, respectively; and $\sigma_{f_i}$ and $\sigma_{f_j}$ are the flux uncertainties.  For our three epochs of observation, we therefore consider source variability over three possible timescales in each band. For a source to be classified as variable over a given timescale, both the [3.6] and [4.5] bands should show variability indices of $|V| > 1.7$ in the same direction (brightening or dimming). This criterion (illustrated in Figure~\ref{fig:M32_varCalc}) was calculated from our data-set assuming a bivariate Gaussian distribution for $V_{3.6\mu{\rm m}}$ and $V_{4.5\mu{\rm m}}$ and corresponds to the 2$\sigma$ joint probability value. This mitigates against photometric mismatches and fluxes enhanced by a nearby or blended source in this high-stellar-density field.

\begin{table}
\centering
 \caption{Number of variable sources detected in M32. If a candidate variable in one epoch is considered to be a high-confidence in another epoch the latter classification is given propriety in Table~\ref{tab:M32catDescription}.}
 \label{tab:IntervalTimescales}
\centering
 \begin{tabular}{@{}lccc@{}}
   \hline
   \hline
 Epochs               &     Interval   &   High-Confidence  &  Candidate  \\
                     &      (days)    &   Variables        &  Variables  \\
\hline  
Epoch 1 - Epoch 2    &       32           &            4                  &        23            \\
Epoch 1 - Epoch 3    &       381          &           14                  &        29            \\
Epoch 2 - Epoch 3    &       349          &           21                  &        43             \\
  \hline
 \end{tabular}
\end{table}


Figure~\ref{fig:M32_varCalc} shows $|V|$ at both 3.6 and 4.5 $\mu$m, for every time interval combination separating the warm {\em Spitzer} data. Stars with a 3$\sigma$ variability index in the 2D Gaussian probability distribution are considered high-confidence variables, whilst stars with a joint probability between 2$\sigma$ and 3$\sigma$ 
are considered candidate variables. This criterion was adopted to enable the recovery of large-amplitude variables which may become fainter than the detection threshold at one or more epochs through their pulsation period.  Our detection criteria resulted in the identification of 28 unique high-confidence variables and 55 candidate variables in M32. 
Table~\ref{tab:IntervalTimescales} lists the time-scales between epochs and the number of variable sources detected. 
If a star is not variable in one pair of epochs it can still be classified as a variable if it has a high variability index in another pair of epochs.
Potentially other candidate variable stars could be identified in M32 using the criteria: $|V_{\rm 3.6}| >1$, and an absolute 3.6 $\mu$m magnitude above the TRGB at 3.6 that \citet{Goldman2019} employed for the Dust in Nearby Galaxies with \emph{Spitzer} (DUSTiNGS) sample of Local Group dwarf galaxies, or by using one of Stetson's indices \citep{Welch1993, Stetson1996}, which \cite{Javadi2011, Javadi2015} and \cite{Saremi2020} have used to identify LPVs in M33 and the Andromeda I dwarf galaxy. However, due to the compact nature of M32 and potential for photometric blends towards the core, we do not adopt these selection criteria for M32.  

Due to sensitivity and the time sampling of our observations, our variability search is biased toward discovering large-amplitude LPVs \citep{Boyer2015b}. 
Infrared-bright populations such as thermally pulsing (TP)-AGB stars are variable on time-scales of 60 to 1000 days \citep{Vassiliadis1993}, with amplitudes up to 2 mag at 3.6 $\mu$m \citep{LeBertre1992,LeBertre1993,McQuinn2007,Vijh2009,Polsdofer2015, Goldman2019, Karambelkar2019}. In
contrast Cepheid variables or RR Lyrae stars, which have shorter periods (typically between 1--80 days), 
are not expected to be recovered as their light-curves are not favourably sampled by our survey.
We are also unlikely to detect small-amplitude ($\Delta m_{3.6} < 0.25$ mag) variable stars, as these will be masked by the photometric errors, especially for sources near the completeness limit. 
Other sources that fluctuate on long cadences include eclipsing binaries and active galactic nuclei (AGN; which tend to be redder than the variables identified in this work). However, their lower amplitudes and irregularity mean they are expected to contribute negligibly to our variable sample \citep{Ulrich1997,Neugebauer1999,Vijh2009,Kozlowski2010, Kozlowski2016, Boyer2015b, Chen2018}.



\begin{figure}
\centering
\includegraphics[trim=0cm 0cm 0cm 0cm, clip=true,width=\columnwidth]{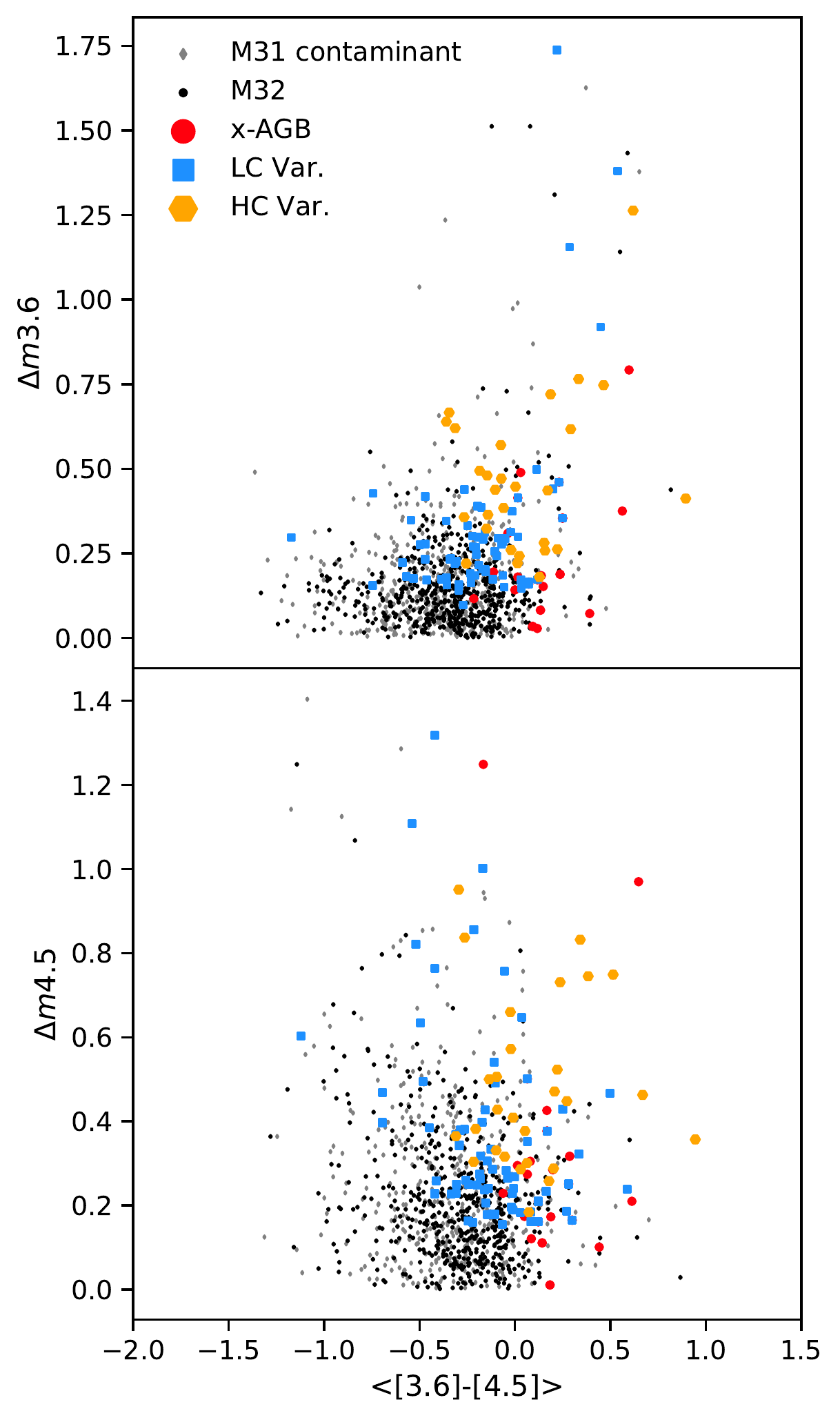}
\caption{Maximum observed amplitude $\Delta m_{3.6}$ (top) and $\Delta m_{4.5}$ (bottom) for each filter compared to the [3.6]--[4.5] colour. Non-variable M32 sources are plotted in black and sources in the M32 field statistically identified as M31 contaminants are in grey. Stars identified as candidate or high confidence variables are highlighted in blue and orange, respectively (see Figure \ref{fig:M32_varCalc}), and extreme AGB stars (see Section~\ref{sec:M32cmds}) not identified as variable in red.} 
  \label{fig:M32_amp}
\end{figure}

In evolved stars, pulsation amplitudes increase as the star evolves along the AGB \citep{Vassiliadis1993, Ita2004, Whitelock2003}. For each source in each band, we compute the difference between the brightest and dimmest magnitudes ($\Delta m_{3.6}$ and $\Delta m_{4.5}$). We plot these against the mean [3.6]--[4.5] colour in Figure~\ref{fig:M32_amp}. The variable stars in our sample show amplitudes in the range $0.2 \lesssim \Delta m \lesssim 2$ mag.  In general, the stars in our sample with the largest $\Delta m_{3.6}$ amplitude variations correspond to the reddest stars (see Section~\ref{sec:M32cmds}). These red sources have a strong mid-IR excess due to dust, which supports existing results linking pulsation strength to dust production among AGB stars \citep[e.g.][]{Whitelock1987, McDonald2018}.

The bottom panel of Figure~\ref{fig:M32_amp} compares $\Delta m_{4.5}$ to the mean [3.6]--[4.5] colour. No systematic increase or decrease in the amplitude with colour is seen. 
While our limited phase coverage means $\Delta m_{4.5}$ recovers only part of the variability amplitude of each source, CO and SiO absorption in the photosphere of cool giants (at 4.08 and 4.66 $\mu$m, respectively; \citealt{Marengo2007}) also has a significant influence on the 4.5 $\mu$m flux; the strength of this absorption changes as the star pulsates, effectively concealing any trend in $\Delta m_{4.5}$ amplitude with colour.
This observed trend is reproduced by the grid of {\sc darwin} models presented in \citet{Bladh2015, Bladh2019}, which produces time-dependent radial structures of the atmospheres and winds of AGB stars allowing us to explore the effects of pulsation (S.~Bladh, priv.~comm.).

\begin{figure} 
\centering
\includegraphics[trim=0cm 0cm 0cm 0cm, clip=true,width=1\columnwidth]{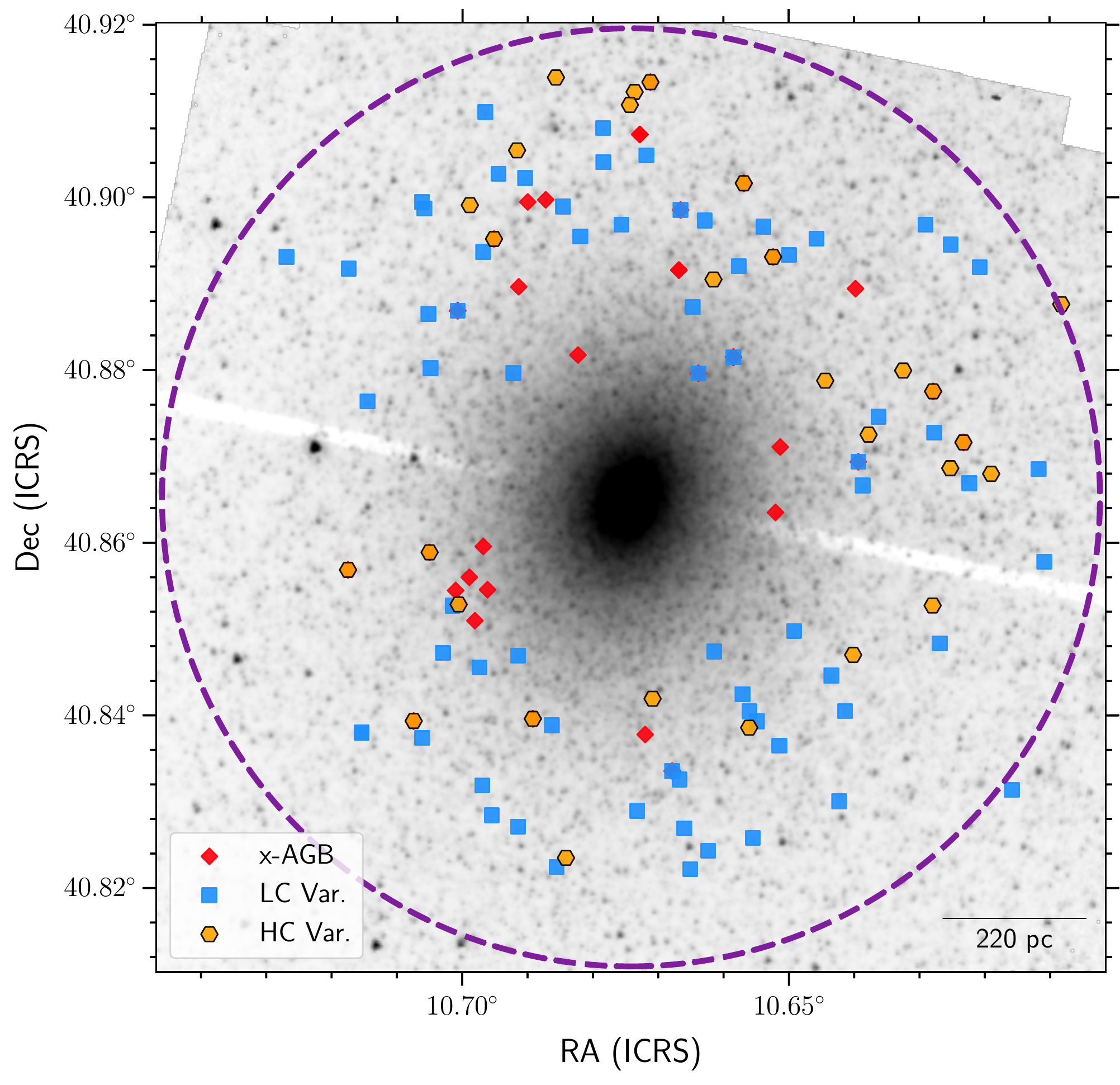}
\caption{The spatial distribution of the candidate variable and x-AGB stars overlaid  on a grey-scale image of M32 at 3.6 $\mu$m. Orange points denote the high-confidence variables, blue squares denote variable-star candidates and red diamonds represent x-AGB stars not identified as variable. The M31 spiral arms and dust lanes are to the north of this image. The dashed purple line indicates the region where M31 begins to dominate the source density. North is up and East is to the left.} 
  \label{fig:M32_varI1}
\end{figure}

Figure~\ref{fig:M32_varI1} shows the spatial distribution of the individual variables over the M32 IRAC 3.6 $\mu$m map. The variable stars are distributed uniformly over the area explored {\em in M32}.
Within our M32 field is the red transient AT 2016hbq \citep{Hornoch2016}; this star was identified as SSTM32-387 by \citet{Jones2015a} and is moderately red at {\em Spitzer} wavelengths with [3.6]--[8.0] = 0.45 mag. Whilst it is detected in our catalogue is not identified as a variable source in our data.  
At larger radii M32's population becomes entangled with the M31 disc populations; from the radial profile in Section~\ref{sec:Completeness},  the mid-IR tidal radius (inferred from photometry by \cite{Jones2015a}) and the {\em WISE} half-light radii computed by \cite{Jarrett2019}, we infer that disc and halo stars from M31 start to dominate at $R \gtrsim 3.4^\prime$. This division between populations is marked in Figure~\ref{fig:M32_varI1}; within this radius we assume the star is a member of M32.

\subsection{The mid-IR stellar populations of M32}
\label{sec:M32cmds}

\begin{figure} 
\centering
\includegraphics[trim=0cm 0cm 0cm 0cm, clip=true,width=\columnwidth]{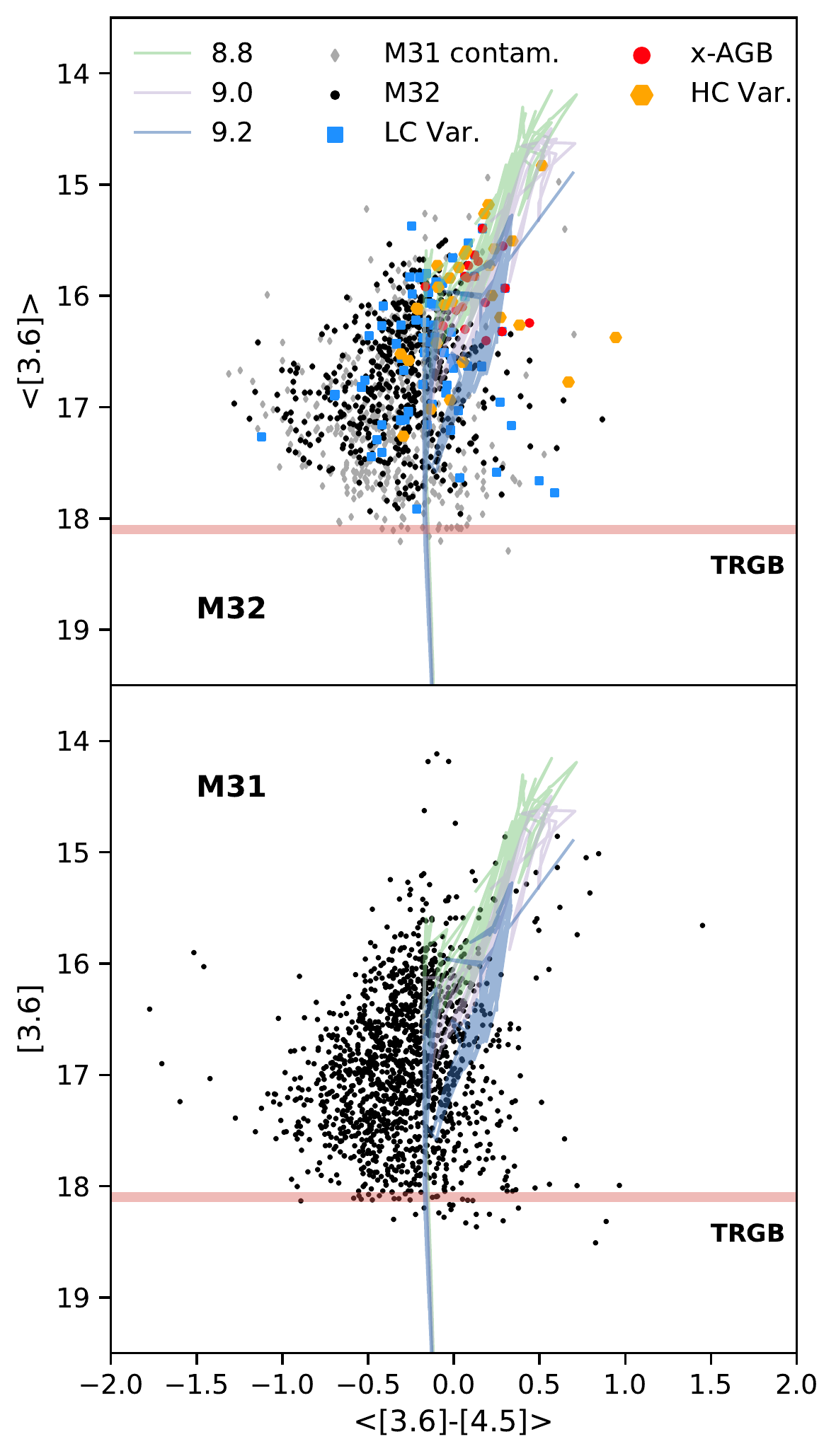}
 \caption{The corrected IRAC $[3.6]-[4.5]$ versus [3.6] CMD, for M32 (top) and the M31 disc field (bottom). The M32 sources detected as variables are shown in blue and orange, and x-AGB stars are shown in red.  Stars in the M32 field statically considered to be M31 contaminants are shown as grey diamonds. The solid pale red lines show the assumed TRGB. Padova isochrones \citep{Marigo2017,Pastorelli2019} 
 for a metallicity of [M/H] = $-$0.11 and ages of log(t/yr) $\sim$ 8.8, 9.0 and 9.2 are shown for reference.}
  \label{fig:M32_343CMD}
\end{figure}

The mean [3.6] versus $[3.6]-[4.5]$ CMDs for M32 and the M31 disc are presented in Figure~\ref{fig:M32_343CMD},  which also highlights the colours of the variable-star candidates compared to the general population of M32. 
Variable stars brighter than the assumed TRGB ($m_{3.6} \sim 18.2$ mag) are classified as TP-AGB candidates, and following \cite{Boyer2015a,Boyer2015b} we identify AGB stars that are likely in the super-wind phase phase of evolution (the candidate extreme AGB; x-AGB stars) as those brighter than $M_{3.6} = -8$ mag and redder than $[3.6]-[4.5] = 0.1$ mag in at least one epoch.  Approximately 60\% of the extreme AGB stars have been detected as a variable. The variable x-AGB sources are expected to have high dust-production rates ($>10^{-7}$ ${\rm M}_{\odot} \, {\rm yr}^{-1}$) and are a subset of the general TP-AGB population \citep{Boyer2011, Jones2017b}. In the Magellanic Clouds, x-AGB stars are typically carbon-rich \citep{vanLoon2006,  Woods2012, Ruffle2015}, with a small fraction (<10\%) associated with dust-enshrouded oxygen-rich AGB stars \citep{Jones2014, Jones2017b}. 
M32 is more metal-rich than the Magellanic Clouds, thus its x-AGB stars can be expected to contain a higher fraction of oxygen-rich stars, due to the higher natal oxygen abundance and subsequent difficulty in achieving C/O $>$ 1, however spectroscopic confirmation is needed to determine the exact ratio between each chemical type in M32.  

M32 is expected to have little star formation in the last 50 Myr \citep{Monachesi2011}, thus massive stars with $M \geq 8$ M$_\odot$ are unlikely to be present in our M32 sample. Any stars brighter than the tip of the AGB ($m_{3.6}$ $\sim 13.7 ~\pm$ 0.6 mag; \citealt{Jones2015a}), are expected to be either luminous RSGs in M31 or foreground sources, due to the short evolutionary time 5--8 \Msun~stars spend in the AGB phase.
Most AGB stars in M32 have slightly blue [3.6]--[4.5] colours typical of oxygen-rich giants, due to photospheric CO and SiO absorption in the 4.5 $\mu$m filter \citep{McQuinn2007, Bolatto2007}. This suggests that the TP-AGB candidates are more likely to be oxygen-rich than carbon-rich, and have little dust emission.  

Carbon stars appear over a limited range of the metallicity-age plane. At approximately solar metallicity, carbon stars are expected to form from stars with initial masses that are about 1.5--5 \Msun \citep{Karakas2007, Marigo2007, Boyer2013, Boyer2019}, and reach the thermal pulsing AGB phase in under 3 Gyr. 
In old populations, bright carbon stars are typically younger than their oxygen-rich counterparts which generally have lower initial masses ($<1.5$\Msun) and start to thermally pulse much later.  However, in intermediate-age populations the brightest AGB variables are O-rich hot bottom burning (HBB) stars \citep[e.g.][]{Menzies2015, Whitelock2018}, with masses between 4--12 \Msun \citep{Doherty2015}.  
Hence, AGB stars trace populations with ages between 13 Gyr to 100 Myr for solar metallicities \citep{Karakas2007, Marigo2007}.

Compared to metal-poor galaxies like the Magellanic Clouds \citep[e.g.][]{Blum2006, Boyer2011}, there is a lack of objects in the M32 CMDs with even moderately red colours, where we would expect carbon-rich AGB stars and objects with significant mass loss to reside. These stars become much rarer at ages exceeding 2 Gy. 
This suggests the population of M32 is skewed older than previously postulated from early {\em Spitzer} data.
To obtain a better estimate of the age of M32's infrared populations we compare our CMDs with theoretical isochrones from \citet{Marigo2017} and \citet{Pastorelli2019}, for a range of stellar ages and metallicity. These {\sc parsec-colibri} stellar isochrones include a detailed treatment of the TP-AGB phase, where several interconnected physical processes affect AGB evolution, e.g. dredge-up, pulsations and dust-driven winds.
Here we consider isochrones with [M/H] of 0, $-$0.11 and $-$0.2 (with solar-scaled abundances) in the age range 0.05--10 Gyr. 
For sources that likely belong to M32, it is shown in Figure~\ref{fig:M32_343CMD}, the data are best represented by moderately metal-poor isochrones with ages between 0.6--3 Gy, and hence expected turn-off masses between 1.5--3 \Msun. 
This is consistent with detailed CMD analyses based on {\em Hubble Space Telescope (HST)} data where it was inferred that the bulk of M32's star formation ceased $\sim$2 Gyr ago \citep{Monachesi2012}. In $[3.6]-[4.5]$ CMDs, stellar sequences from small populations of stars are difficult to isolate, and a more robust comparison of the populations of cool, evolved stars is not possible.


The differences between the stellar populations of M32 and the disc of M31 are revealed in Figure~\ref{fig:M32_343CMD}. 
In general, the two sets of CMDs have a similar morphology regarding their evolved stellar populations. 
The M31 disc pointing has a population of brighter stars at [3.6]--[4.5] = $-$0.3 mag, and [3.6] $<$ 15.5 mag which is completely absent in the M32 field. The majority of these sources are likely to be RSG stars, with a few massive ($M > 4$ \Msun) AGB stars experiencing hot-bottom burning. 
This indicates that no recent star formation has occurred in M32.  A lack of star formation is consistent with M32's observed gas and interstellar dust deficiency \citep{Sage1998}, where most of the gas was likely stripped through interaction with M31's tidal field \citep{Bekki2007}. 
Conversely, the M31 disc is still undergoing star formation, resulting in the massive population of RSGs, seen in the M31 field but not in the contamination-corrected field of M32. 
These results are consistent with the numerous spectroscopic and photometric studies that conclude stars in M32 are older than the stellar populations in our M31 disc field. 

Note that we have assumed that the population of M32 is negligible in the M31 disc field, and that the M31 field is fully representative of the M31 population underlying M32; if there were substantial M32 contamination in this region or a comparative surplus of M31 stars in this region then we have over-corrected the M32 CMD resulting in missing population information in our primary field, however this is unlikely \citep[see e.g.][]{Monachesi2012}.  Furthermore, this comparison between population is only valid for the outer regions of M32 where the level of crowding and photometric incompleteness is negligible compared to the core region where significant crowding prevents us resolving individual stars.


\subsection{Dust-production rates}

AGB dust production and the strength of their pulsations are closely linked \citep{LagadecZijlstra2008, Mcdonald2019}, with a super-wind phase occurring towards the end of their evolution once the star pulsates on periods of $\sim$300 days \citep{Vassiliadis1993}. 
In the late stages of evolution, AGB stars experience intense mass loss, the rate of which can be characterised by using mid-IR colours as a proxy for dust-production rates. 
We use the average [3.6]--[4.5] colour to estimate the dust-production rates of the variable-star candidates using the relation 
\begin{equation}
\log \dot{D} \, [{\rm M}_{\odot} \, {\rm yr}^{-1}] = -9.5+[1.4 \times ([3.6]-[4.5])],  
\end{equation}
adopted by \cite{Boyer2015b} for the DUSTiNGS galaxies. 
Assuming all the TP-AGB stars are carbon rich, we find a  lower limit to the cumulative AGB dust input of 6.4 $\times 10^{-8}$ ${\rm M}_{\odot} \, {\rm yr}^{-1}$. 

 Computing dust-production rates from infrared colour relations relies upon a large number of assumptions. For instance, we assume a single dust composition, wind speed and effective stellar temperature. For individual stars the [3.6]--[4.5] colour may also be contaminated by molecular tracers, and for the same [3.6]-[4.5] colour, oxygen-rich stars have a higher dust-production rate compared to their carbon-rich counterparts, due to the difference in opacity between silicate and amorphous carbon grains. Thus mass-loss rates computed for individual AGB candidates have a large associated uncertainty.

\cite{DellAgli2018, DellAgli2019} found that the correlation between the dust-production rate versus [3.6]--[4.5] colour has a stronger dependence on the underlying properties of the AGB population than the [3.6]--[8.0] colour. 
To account for this, and given that the most extreme stars dominate the dust-production in Local Group Galaxies \citep{Riebel2012, Srinivasan2016, Jones2018a}, we match our candidate variable-stars  to the cold {\em Spitzer} data from \cite{Jones2015a}. Using only variable stars which have an IR-excess of $[3.6] - [8.0] > 0.5$ mag in the cold {\em Spitzer} data, we compute the cumulative dust-production rate using the $[3.6]-[8.0]$ mass-loss-rate--colour relation derived empirically by \cite{Matsuura2009},  which assumes gas-to-dust ratio of 200. This results in a cumulative dust-production rate of 1.1 $\times 10^{-7}$ ${\rm M}_{\odot} \, {\rm yr}^{-1}$.
This should be considered a more reliable estimate of the cumulative dust-return for long-period variable stars in M32, but with only three epochs, aliasing effects and photometric incompleteness will mask a certain fraction of variable sources in M32 and hence this cumulative dust return should be considered a minimum value for the dust-input-rate to this galaxy. 

Like M32, NGC 185, and NGC 147 are dwarf ellipticals galaxies in the local group which have been observed during the warm {\em Spitzer} mission \citep{Boyer2015a, Boyer2015b}. 73 and 94 IR variable star candidates have been detected in NGC 185 and NGC 147, respectively. In both instances the crowded inner regions of these galaxies limited the identification of AGB candidates.
Whilst the number of variable AGB star candidates detected is comparable to M32, the mean [3.6]--[4.5] colour for their AGB populations are redder and thus their cumulative dust-productions rates are estimated to be slightly higher, with both galaxies producing dust at rates of 2.2 -- 2.7 $\times 10^{-7}$ ${\rm M}_{\odot} \, {\rm yr}^{-1}$. 
As all these values are lower limits, it is likely that this difference is due to limitations of the data, and stochastic variation in the number of very red evolved stars identified (which are thought to dominate the dust production in galaxies) rather than due to the physical properties of the host galaxies.




\section{Summary and Conclusions}
\label{sec:conclusion}

In this paper, we presented multi-epoch {\em Spitzer} IRAC observations of the dwarf elliptical galaxy M32. 
We find 28 high-confidence and 55 candidate infrared variable stars. 
We identified and characterised these large-amplitude variables using three epochs of warm {\em Spitzer} data spanning a total of 381 days. 
The variable-source population is dominated by evolved stars, with 20\% classified as extreme AGBs according to their [3.6]--[4.5] colour. 
The evolved stellar population in the (contaminant) M31 and (target) M32 fields have a similar morphology in their [3.6]--[4.5] CMDs, however, an additional population of bright stars thought to be RSG or massive ($M > 4$ \Msun) AGB stars experiencing hot-bottom burning is seen in the M31 disc pointing but not in M32.

Unfortunately, within 1.5' of the M32 core, the severe crowding prohibits us resolving individual stars and at fainter magnitudes the data become too noisy to detect variability down to the RGB tip due to the high surface brightness. 
Looking into the future, the superior angular resolution and sensitivity of the upcoming {\em James Webb Space Telescope} will be able to detect individual stars in much greater depth and closer to the central core of M32, which may help to provide greater clarity on the origins of this enigmatic galaxy.

\section*{Acknowledgements}

We would like to thank Patricia Whitelock for helpful comments and suggestions in the course of this work, Sara Bladh for calculating {\em Spitzer} photometry for the grid of {\sc darwin} models \citep{Bladh2015} and Jacco van Loon for the detailed and constructive review that helped us to improve this paper.
This work is based on observations made with the \spitzer Space Telescope, which is operated by the Jet Propulsion Laboratory, California Institute of Technology under a contract with NASA.
OCJ has received funding from the EUs Horizon 2020 programme under the Marie Sk\l{}odowska-Curie grant agreement No 665593 awarded to the STFC. IM acknowledges support from the UK Science
and Technology Facility Council under grant ST/L000768/1. This research made use of Astropy,\footnote{http://www.astropy.org} a community-developed core Python package for Astronomy \citep{Astropy2013}.

\

\noindent {\it Facilities:}  {\em Spitzer} (IRAC).

\

\noindent The data underlying this article are available in the article and in its online supplementary material.




\def\aj{AJ}					
\def\actaa{Acta Astron.}                        
\def\araa{ARA\&A}				
\def\apj{ApJ}					
\def\apjl{ApJL}					
\def\apjs{ApJS}					
\def\ao{Appl.~Opt.}				
\def\apss{Ap\&SS}				
\def\aap{A\&A}					
\def\aapr{A\&A~Rev.}				
\def\aaps{A\&AS}				
\def\azh{AZh}					
\def\baas{BAAS}					
\def\jrasc{JRASC}				
\def\memras{MmRAS}				
\def\mnras{MNRAS}				
\def\pra{Phys.~Rev.~A}				
\def\prb{Phys.~Rev.~B}				
\def\prc{Phys.~Rev.~C}				
\def\prd{Phys.~Rev.~D}				
\def\pre{Phys.~Rev.~E}				
\def\prl{Phys.~Rev.~Lett.}			
\def\pasp{PASP}					
\def\pasj{PASJ}					
\def\qjras{QJRAS}				
\def\skytel{S\&T}				
\def\solphys{Sol.~Phys.}			
\def\sovast{Soviet~Ast.}			
\def\ssr{Space~Sci.~Rev.}			
\def\zap{ZAp}					
\def\nat{Nature}				
\def\iaucirc{IAU~Circ.}				
\def\aplett{Astrophys.~Lett.}			
\def\apspr{Astrophys.~Space~Phys.~Res.}		
\def\bain{Bull.~Astron.~Inst.~Netherlands}	
\def\fcp{Fund.~Cosmic~Phys.}			
\def\gca{Geochim.~Cosmochim.~Acta}		
\def\grl{Geophys.~Res.~Lett.}			
\def\jcp{J.~Chem.~Phys.}			
\def\jgr{J.~Geophys.~Res.}			
\def\jqsrt{J.~Quant.~Spec.~Radiat.~Transf.}	
\def\memsai{Mem.~Soc.~Astron.~Italiana}		
\def\nphysa{Nucl.~Phys.~A}			
\def\physrep{Phys.~Rep.}			
\def\physscr{Phys.~Scr}				
\def\planss{Planet.~Space~Sci.}			
\def\procspie{Proc.~SPIE}			
\let\astap=\aap
\let\apjlett=\apjl
\let\apjsupp=\apjs
\let\applopt=\ao


\bibliographystyle{mnras}
\bibliography{libby} 







\bsp	
\label{lastpage}
\end{document}